\documentclass{jfm}
\usepackage{graphicx}
\usepackage{epstopdf, epsfig}
\usepackage{empheq}
\usepackage{url}

\shorttitle{The three-dimensional structure of swirl-switching in bent pipe flow}
\shortauthor{L.~Hufnagel, J.~Canton, R.~\"{O}rl\"{u}, O.~Marin, E.~Merzari and P.~Schlatter}

\title{The three-dimensional structure of swirl-switching in bent pipe flow}

\author{%
	Lorenz Hufnagel\aff{1},
	Jacopo Canton\aff{1}\corresp{\email{jcanton@mech.kth.se}},
	Ramis \"{O}rl\"{u}\aff{1},
	Oana Marin\aff{2},
	Elia Merzari\aff{2}
	\and Philipp Schlatter\aff{1}
}

\affiliation{%
	\aff{1}
	Linn\'e FLOW Centre and Swedish e-Science Research Centre (SeRC),
	KTH Mechanics,
	Royal Institute of Technology,
	Stockholm, SE-100 44, Sweden
	\aff{2}
	Mathematics and Computer Science Division,
	Argonne National Laboratory,
	Argonne, IL, USA
}

\newcommand{\ie}{\emph{i.e., }}

\begin{document}

\maketitle

%-------------------------------------------------------------------------------
\begin{abstract}
	Swirl-switching is a low-frequency oscillatory phenomenon which affects the
	Dean vortices in bent pipes and may cause fatigue in piping systems.
	Despite thirty years worth of research, the mechanism that causes these
	oscillations and the frequencies that characterise them remain unclear.
	Here we show that a three-dimensional wave-like structure
	is responsible for the low-frequency switching of the dominant Dean vortex.
	The present study, performed via direct numerical simulation,
	focuses on the turbulent flow through a $90^\circ$ pipe bend preceded and
	followed by straight pipe segments.
	A pipe with curvature $0.3$ (defined as ratio between pipe radius and bend
	radius) is studied for a bulk Reynolds number $\Rey =	11\,700$,
	corresponding to a friction Reynolds number $\Rey_\tau\approx360$.
	Synthetic turbulence is generated at the inflow section and used instead of
	the classical recycling method in order to avoid the interference between
	recycling and swirl-switching frequencies.
	The flow field is analysed by three-dimensional proper orthogonal
	decomposition (POD) which for the first time allows the identification of
	the source of swirl-switching: a wave-like structure that originates in the pipe
	bend.
	Contrary to some previous studies, the flow in the upstream pipe does not
	show any direct influence on the swirl-switching modes.
	Our analysis further shows that a three-dimensional characterisation of the
	modes is crucial to understand the mechanism, and that reconstructions based
	on 2D POD modes are incomplete.
\end{abstract}


%-------------------------------------------------------------------------------
\section{Introduction}
\label{sec:intro}
\begin{figure}
	\centering
	\includegraphics{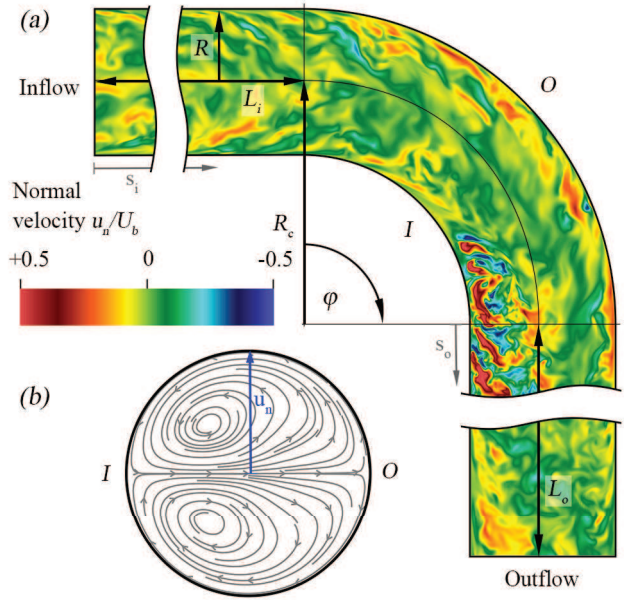}
	\caption{%
		Schematic of the computational domain.
		(a) A section of the pipe with curvature $\delta = R/R_c = 0.3$,
		including the definition of the geometrical parameters and an
		instantaneous flow field coloured by normal velocity, \ie normal to the
		bend symmetry plane.
		(b) Streamlines of the mean cross-flow showing the Dean
		vortices on a cross-section extracted at $s_o=0$.
		An animated view of the setup is provided in the supplementary online material, \texttt{movie1.m4v}.
	}
	\label{fig:domain_sketch}
\end{figure}

Bent pipes are an essential component of a large number of industrial machines
and processes.
They are ideal for increasing mass and momentum transfer, passively mixing
different fluids, which makes them effective as heat exchangers, inverters, and
other appliances.
For a review of the applications of bent pipes in industry see
\citet{Vashisth2008a}; the most recent advances in experiments and
simulations can be found in the review by \citet{KalpakliVester2016}.
The high mass and momentum transfer is generated by the secondary motion
caused by the centrifugal force acting on the fluid in the curved sections.
This secondary motion, which is of Prandtl's first kind, takes the shape of two
counter-rotating vortices, illustrated in figure~\ref{fig:domain_sketch}, which
move the fluid towards the outside of the bend, along the centreline, and back
towards the inside along the wall, therefore increasing the mass and momentum
transfer across the pipe section.

These vortices were first observed by \citet{Boussinesq1868} and
\citet{Eustice1910}, and later described analytically by \citet{Dean1928a} from
whom they received the name of Dean vortices.
The intensity of these vortices increases with Reynolds number, here based on
pipe diameter and bulk velocity (\ie $\Rey = D U_b / \nu$, where $\nu$ is the
kinematic viscosity of the fluid), as well as with pipe curvature, defined as
the ratio between pipe radius and bend radius, $\delta = R/R_c$
(see \citealt{Canton2016b} for laminar flow; \citealt{Noorani2013}, the
review by \citealt{KalpakliVester2016}, and references therein, for turbulent
flows).

For laminar, steady flow the Dean vortices are symmetric with respect to the
bend symmetry plane (the $I$-$O$ plane in figure~\ref{fig:domain_sketch}); but
when the flow becomes unstable the vortices start oscillating periodically
\citep{Kuhnen2014,Kuhnen2015,Canton2016a}.
These large-scale oscillations are caused by the appearance of periodic
travelling waves which, as also observed in other flows
\citep[see, \textit{e.g.}][]{Hof2004}, are at the base of transition to
turbulence for toroidal and helical pipes.

A different kind of large-scale oscillations is observed for high Reynolds
numbers: here the turbulent flow is modulated by a low-frequency alternation of
the dominant Dean vortex.
This vortex alternation excites the pipe structure and is presumed to be the
cause of structural, low-frequency oscillations observed in heat exchangers
\citep[\textit{e.g.} in microgravity conditions such as in a test for the international space
station;][]{Brucker1998}, as well as the origin of secondary motion in the bends
of the cooling system of nuclear reactors \citep{KalpakliVester2016}.
The Dean vortex alternation was initially, and
unexpectedly, observed by \citet{Tunstall1968}, who experimentally studied the
turbulent flow through a sharp, L-shaped bend ($\delta = 1$).
These authors measured ``low random-frequency'' switches between two distinct
states and, by means of flow visualisations, were able to identify an either
clockwise or anti-clockwise predominance of the swirling flow following the
bent section.
The switching was found to have a Strouhal number $St = f D/U_b$ highly
dependent on $\Rey$ and comprised between $2\times10^{-4}$ and
$4.5\times10^{-3}$.
Tunstall \& Harvey attributed the origin of the switching to the presence of a
separation bubble in the bend and to the ``occasional existence of
turbulent circulation entering the bend''.

To the best of our knowledge, the first author to continue the work by
\citet{Tunstall1968} was \citet{Brucker1998}, who analysed the phenomenon via
particle image velocimetry (PIV) and coined the term ``swirl-switching''.
Br\"ucker studied a smoothly curved pipe with $\delta = 0.5$ and
identified the oscillations as a continuous transition between
two mirror-symmetric states with one Dean cell larger than the other.
He confirmed that the switching takes place only when the flow is turbulent, and
he reported two distinct peaks, $St = 0.03$ and $0.12$, at frequencies
considerably higher than those measured by Tunstall and Harvey, despite the
lower Reynolds numbers considered.

\citet{Rutten2001,Rutten2005} were the first to numerically study
swirl-switching by performing large-eddy simulations (LES) for $\delta = 0.167$
and $0.5$.
The main result of this analysis is that the
switching takes place even without flow separation; moreover, R\"utten
and co-workers
found that the structure of the switching is more complex than just
the alternation between two distinct symmetric states, since the outer
stagnation point ``can be found at any angular position within $\pm40^\circ$''.
R\"utten \etal\ found a high-frequency peak at $St\approx0.2$, attributed to
a shear-layer instability, and, only for their high Reynolds number case,
one low-frequency peak for $St\approx5.5\times10^{-3}$, which was
connected to the swirl-switching.
However, the simulations for R\"utten \etal's work were performed by using a
``recycling'' method, where the results from a straight pipe simulation were used
as inflow condition for the bent pipe.
These periodic straight pipes were of length $L=3.5D$ and $5D$ and likely
influenced the frequencies measured in the bent pipes since the periodicity
of the straight pipes
introduced a forcing for $St = U_b / L = 1/3.5$ and $1/5$, respectively.

\citet{Sakakibara2010} were the first to analyse the flow by means of
two-dimensional proper orthogonal decomposition (2D POD) performed on snapshots
extracted from stereo PIV.
Their results for $\delta=0.75$ reveal anti-symmetric structures that
span the entire pipe cross-section and contain most of the energy of the flow.
A spectral analysis of the corresponding time coefficients shows peaks between
$St \approx 0.07$ at $s_o = 2D$ and $St \approx 0.02$ for $s_o = 25D$, in the
range found by \citet{Brucker1998}.
In a subsequent work \citet{Sakakibara2012} conjectured that the
swirl-switching is caused by very large-scale motions (VLSM) formed in the
straight pipe preceding the bend.

\citet{Hellstrom2013} and \citet{KalpakliVester2013} also presented
results based on 2D POD.
The former performed experiments for $\delta = 0.5$ and found non-symmetric
modes resembling a tilted variant of the Dean vortices with $St = 0.16$ and
$0.33$, corresponding to the shear-layer instabilities found by
\citet{Rutten2005}.
\citet{KalpakliVester2013}, on the other hand, studied a pipe with
$\delta=0.31$; differently from previous works, the section of straight
pipe following the bend was only $0.67$ diameters long.
Their results at the exit of this short segment show clearly
antisymmetric modes as most dominant structures.
The swirl-switching frequency obtained from the POD time coefficients was
$St=0.04$; peaks of $St = 0.12$ and $0.18$ were also measured but were found
not to be related to swirl-switching.
In a later work \citet{KalpakliVester2015} repeated the experiments for
$\delta=0.39$ and found again a dominant frequency corresponding to $St=0.04$.

\citet{Carlsson2015} performed LES in a geometry similar to that of
\citet{KalpakliVester2013}, namely, with a short straight section following the
bend, for four different curvatures.
The inflow boundary condition was generated by means of a recycling method, as in
\citet{Rutten2001,Rutten2005}, with a straight pipe of length $7D$, exciting
the flow in the bent pipe at $St = 1/7$.
The three lower curvatures were therefore dominated by the spurious frequencies
artificially created in the straight pipe by the
recycling method, while the frequencies measured for $\delta=1$,
corresponding to $0.5 < St < 0.6$, were in the same range identified by
\citet{Hellstrom2013} but were found to be mesh dependent.

\citet{Noorani2016} were the first to investigate the swirl-switching by means
of direct numerical simulations (DNS).
By using a toroidal pipe they showed that swirl-switching is not
caused by structures
coming from the straight pipe preceding the bend, but is a phenomenon inherent
to the curved section.
Two curvatures were investigated, $\delta=0.1$ and $0.3$, and both presented a
pair of antisymmetric Dean vortices as the most energetic POD mode with
$St = 0.01$ and $0.087$.

Table~\ref{tab:literature_strouhal} summarises the main results of the
aforementioned studies.
It is clear from this literature review that there is a strong disagreement
among previous works not only on what is the mechanism that leads to
swirl-switching, but also on what is the frequency that characterises this
phenomenon.
In the present work an answer to both questions will be given, which will also
explain the discrepancies between previous studies.

The paper continues with a description of the numerical methods employed for the
analysis, presented in \S\ref{sec:methods}, devoting special attention
to the inflow boundary conditions.
The results of the simulations and POD analysis are presented in
\S\ref{sec:results} and are discussed and compared
with the literature in \S\ref{sec:conclusions}.

\begin{table}
	\centering
	\begin{tabular}{lccc}
		Reference & $\delta$ & $\Rey$ & $St$
		\\[3pt]
		\citet{Tunstall1968}          & $1$                       & $50\,000$ -- $230\,000$ & $2\times10^{-4}$ -- $4.5\times10^{-3}$   \\
		\citet{Brucker1998}           & $0.5$                     & $5\,000$               & $0.03$, $0.12$                          \\
		\citet{Rutten2001,Rutten2005} & $0.167$, $0.5$            & $27\,000$              & $5.5\times10^{-3}$                      \\
		\citet{Sakakibara2010}        & $0.75$                    & $120\,000$             & $0.02$ -- $0.07$                         \\
		\citet{Hellstrom2013}         & $0.5$                     & $25\,000$              & $0.16$, $0.33$                          \\
		\citet{KalpakliVester2013}    & $0.31$                    & $34\,000$              & $0.04$                                  \\
		\citet{KalpakliVester2015}    & $0.39$                    & $24\,000$              & $0.04$                                  \\
		\citet{Carlsson2015}          & $0.32$, $0.5$, $0.7$, $1$ & $34\,000$              & $0.003$ -- $0.01$, $0.13$, $0.5$ -- $0.6$ \\
		\citet{Noorani2016}           & $0.1$, $0.3$              & $11\,700$              & $0.01$, $0.087$                         \\
	\end{tabular}
	\caption{%
    Reference Strouhal numbers measured in previous studies and attributed to
    swirl-switching.
    The analysis in the present work is performed at $\Rey=11\,700$ in a
    bent pipe with curvature $\delta=0.3$.
	}
	\label{tab:literature_strouhal}
\end{table}

%-------------------------------------------------------------------------------
\section{Analysis methods}
\label{sec:methods}

%-------------------------------------------------------------------------------
\subsection{Numerical discretisation}
\label{subsec:discretisation}
The present analysis is performed via DNS of the incompressible Navier--Stokes
equations.
The equations are discretised with the spectral-element code \texttt{Nek5000}
\citep{nek5000} using a $\mathbb{P}_N-\mathbb{P}_{N-2}$ formulation.
After an initial mesh-dependency study, the polynomial order was set to $N=8$
for the velocity and, consequently, $N=6$ for the pressure.
We consider a $90^\circ$ bent pipe with
curvature $\delta=0.3$ for a Reynolds number 
$\Rey=11\,700$, corresponding to a friction Reynolds number $\Rey_\tau \approx 360$ 
(referred to the straight pipe sections).
A straight pipe of length $L_i = 7D$ precedes 
the bent section (see \S\ref{subsec:syem_validation}), and a second straight 
segment of length $L_o = 15D$ follows it.
Further details about the mesh, including element number and size, are
reported in table~\ref{tab:mesh}.
The supplementary video \texttt{movie1.m4v} shows the setup and a
visualization of the flow.

\begin{table}
	\centering
	\begin{tabular}{lccccc}
		~            & $n_{\mathrm{el}}$ & $n_{\mathrm{dof}}$ & $\Delta r^+$   & $\Delta (R\theta)^+$ & $\Delta z^+$ \\[3pt]
   	$\delta=0.3$ & $480\,000$        & $595\,258\,500$    & $(0.56, 4.89)$ & $ (2.26, 4.40)$      & $(0.93, 10.29)$    \\
	\end{tabular}
	\caption{%
    Details of the mesh employed in the present work.
		$n_{\mathrm{el}}$ corresponds to the number of hexahedral elements, while
		$n_{\mathrm{dof}}$ is the total number of degrees of freedom (velocity
		and pressure).
		Quantities indicated as $(\cdot)^+$ are reported in viscous scaling, and
		the numbers between parenthesis correspond to minimum and maximum values.
	}
	\label{tab:mesh}
\end{table}

%-------------------------------------------------------------------------------
\subsection{Inflow boundary and divergence-free synthetic eddy method}
\label{subsec:syem}
Since the aim of the present work is to reproduce and study swirl-switching, a
periodic or quasi-periodic phenomenon, the treatment of the inflow boundary is
of utmost importance.
The flow field prescribed at the inflow boundary should not introduce any
artificial frequency, in order to avoid the excitation of unphysical phenomena
or a modification of the frequencies inherent to the swirl-switching.
A recycling method, as the one used by \citet{Rutten2001,Rutten2005} and
\citet{Carlsson2015}, should therefore be avoided, as highlighted in
\S\ref{sec:intro}.

In the present work the velocity field at the inlet boundary of the straight
pipe preceding the bend is prescribed via a divergence-free synthetic eddy
method (DFSEM).
This method, introduced by \citet{Poletto2011} and based on the original work
by \citet{Jarrin2006}, works by prescribing a mean flow modulated in time by
fluctuations in the vorticity field.
The superposition of the two reproduces up to second order the mean turbulent
fluctuations of a reference flow and requires a short streamwise adjustment
length to fully reproduce all quantities.
The fluctuations are provided by a large number of randomly distributed
``vorticity spheres'' (or ``eddies'') which are generated and advected
with the bulk velocity in a fictitious cylindrical container located
around the inflow section.
When a sphere exits the container, a new, randomly located sphere is created to
substitute it.
The cylindrical container is dimensioned such that newly created eddies 
do not touch the inlet plane upon their creation and they have stopped
affecting it before exiting the container, \ie the cylinder extends from
$-\max(D_{\text{eddies}})$ to $\max(D_{\text{eddies}})$
\citep[see figure~5 in][for an illustration of the container]{Poletto2013}.

The random numbers required to create the fluctuations are generated on a
single processor with a pseudo random number generator \citep{Chandler2003}
featuring an algorithmic period of $2^{1376}$ iterations, large enough to
exclude any periodicity in the simulations, which feature approximately
$10\,000$ synthetic eddies.
The size of the spheres is selected to match the local integral
turbulence length scale, and their intensity is scaled to recover the
reference turbulent kinetic energy, producing isotropic but heterogeneous
second-order moments.
The method prescribes isotropic turbulence, instead of the correct anisotropic
variant, because it was shown that the former leads to a shorter adjustment
length in wall-bounded flows \citep[see figure~11 in][]{Jarrin2006}.
In order to satisfy the continuity equation, no synthetic turbulence is created
below $(R-r)^+ < 10$; however, this does not significantly affect the
adjustment length since the dynamics of the viscous sublayer
are faster than the mean and converge to a fully developed state
in a shorter distance.
The turbulence statistics necessary for the method, specifically the mean flow
$U(r)$, the turbulent kinetic energy $k(r)$, and the dissipation rate
$\epsilon(r)$, were extracted from the straight pipe DNS performed by
\citet{ElKhoury2013}.
Section~\ref{subsec:syem_validation} presents the validation of our
implementation of the DFSEM; more details can be found in \citet{Hufnagel2016}.

%-------------------------------------------------------------------------------
\subsection{Proper orthogonal decomposition}
\label{subsec:pod}
Besides point measures, we use POD \citep{Lumley1967} to extract coherent
structures from the DNS flow fields and identify the mechanism responsible for
swirl-switching.
More specifically, we use snapshot POD \citep{Sirovich1987} where $n$
three-dimensional, full-domain flow fields
of dimension $d$ (corresponding to the number of velocity unknowns) are stored as snapshots.
POD decomposes the flow in a set of orthogonal spatial modes
$\boldsymbol{\varPhi}_i(\boldsymbol{x})$ and corresponding time coefficients
$a_i(t)$ ranked by kinetic energy content, in decreasing order.
The most energetic structure extracted by POD corresponds to the mean flow and
will be herein named ``zeroth mode'', while the term ``first mode'' will be
reserved for the first time-dependent structure.

A series of instantaneous flow fields (snapshots) is ordered column-wise in a
matrix $\boldsymbol{S} \in \mathbb{R}^{d \times n}$ and decomposed as:
\begin{equation}
	\boldsymbol{S} = \boldsymbol{U} \boldsymbol{\Sigma} \boldsymbol{V}^\intercal
	               = \sum_{i=1}^{d} \boldsymbol{\varPhi}_i a_i,
	\label{eq:pod}
\end{equation}
where
$\boldsymbol{U} \in \mathbb{R}^{d \times d}$,
$\boldsymbol{\Sigma} = diag(\sigma_1,\sigma_2,\cdots,\sigma_m,0)$,
with $m=\min (d,n)$,
and $\boldsymbol{V} \in \mathbb{R}^{n \times n}$.
The decomposition in \eqref{eq:pod} is obtained by computing the singular value
decomposition (SVD) of
$\boldsymbol{M}^{1/2}\boldsymbol{S}\boldsymbol{T}^{1/2}$,
where $\boldsymbol{M}$ is the mass matrix and $\boldsymbol{T}$ is the temporal
weights matrix, which results in
$\tilde{\boldsymbol{U}} \boldsymbol{\Sigma} \tilde{\boldsymbol{V}}^\intercal$,
where $\tilde{\boldsymbol{U}}$ and $\tilde{\boldsymbol{V}}$ are unitary matrices
($\tilde{\boldsymbol{U}}^\intercal\tilde{\boldsymbol{U}} = \boldsymbol{I}$ and
$\tilde{\boldsymbol{V}}^\intercal\tilde{\boldsymbol{V}} = \boldsymbol{I}$);
the POD modes are then obtained as
$\boldsymbol{U} = \boldsymbol{M}^{-1/2}\tilde{\boldsymbol{U}}$ and
$\boldsymbol{V} = \boldsymbol{T}^{-1/2}\tilde{\boldsymbol{V}}$.
To improve the convergence of the decomposition, we exploit the symmetry of the
pipe about the $I-O$ plane, which results into a statistical symmetry for the
flow, and store an additional mirror image for each snapshot
\citep{Berkooz1993}.

%-------------------------------------------------------------------------------
\section{Results and analysis}
\label{sec:results}

%-------------------------------------------------------------------------------
\subsection{Inflow validation}
\label{subsec:syem_validation}

\begin{figure}
	\centering
	\includegraphics[width=\textwidth]{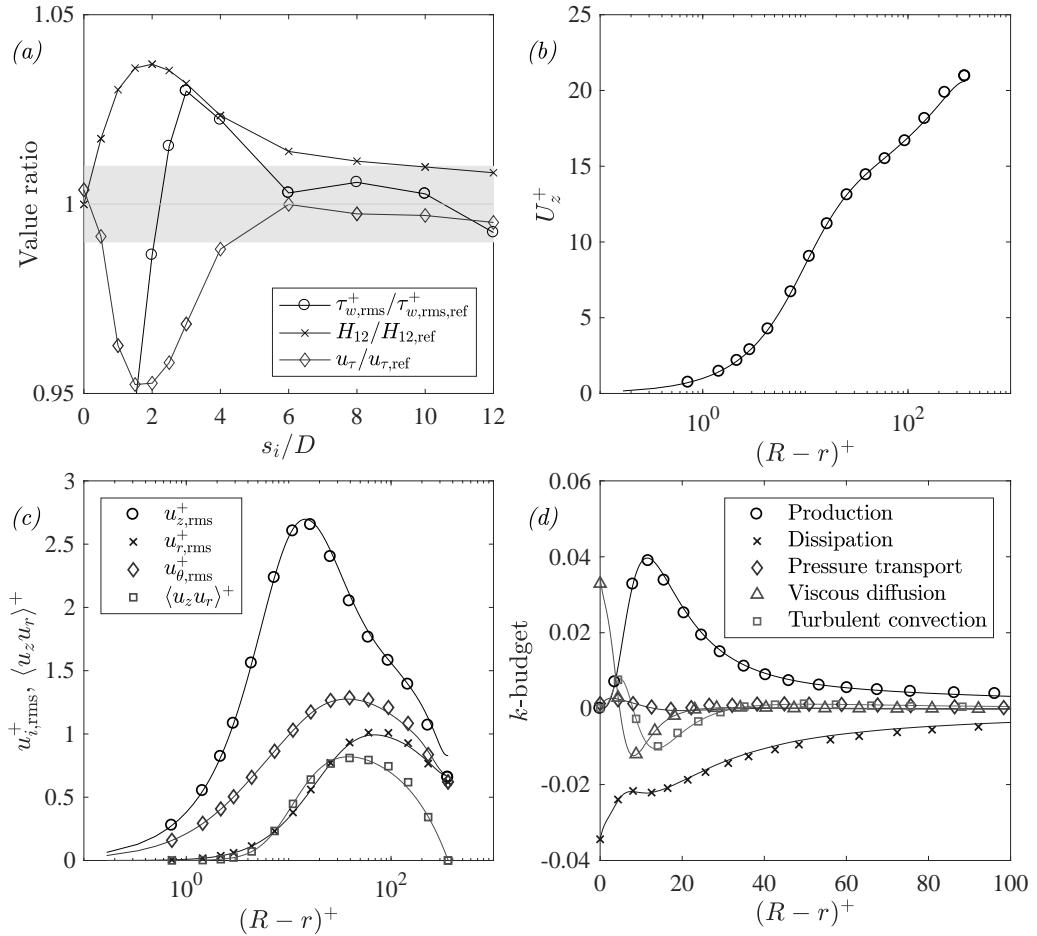}
	\caption{%
		Recovery of fully developed turbulence statistics for the divergence-free
		synthetic eddy method at $\Rey_\tau = 360$, compared to the reference
		values by \citet{ElKhoury2013}.
		Panel \textit{(a)} shows the ratio between the DFSEM and the reference
		data as a function of streamwise distance from the inflow plane.
		The grey shaded area indicates a $\pm 1\%$ tolerance with respect to the
		data by \citet{ElKhoury2013}.
		Panels \textit{(b-d)} show classical statistical profiles as a function of
		radial position at $s_i = 5D$.
		Solid lines indicate the reference data, while symbols represent the
		current results (note that the number of shown points is reduced and does
		therefore not represent the grid resolution; see table~\ref{tab:mesh}).
	}
	\label{fig:integral_q}
\end{figure}

An auxiliary simulation for $\Rey_\tau=360$ was set up to test the performance
of the DFSEM in a $25D$ long straight pipe, provided with the same mesh
characteristics used for the bent pipes.
Classical statistical quantities were used for the validation and
compared with the reference values by \citet{ElKhoury2013}.
The comparison is presented in figure~\ref{fig:integral_q}\textit{(a)} as
a function of distance from the inflow boundary, and shows that
the DFSEM approaches a fully developed turbulent state (within $\pm 1\%$ of
error) at approximately $5D$ from the inflow boundary.
Figures~\ref{fig:integral_q}\textit{(b-d)} present the velocity, stress
profiles, and the turbulent kinetic energy budget at the chosen streamwise
position of $s_i = 5D$, which confirm the recovery of fully developed
turbulence by the divergence-free synthetic eddy method.

A length of $7D$ was therefore chosen for the straight pipe preceding the bent
section, in order to allow for some tolerance and to account for the (weak, up
to $1D$) upstream influence of the Dean vortices \citep{Anwer1989,Sudo1998}.
For comparison, the more commonly used approach where random noise is
prescribed at the inflow requires a development length between $50$ and $110D$
\citep{Doherty2007}.
POD modes were also computed to further check the correctness of this method.
The results, not reported here for conciseness, were in good agreement with
those presented by \citet{Carlsson2015} for a periodic straight pipe, that is,
streamwise invariant modes with azimuthal wavenumbers between $3$ and $7$.

%-------------------------------------------------------------------------------
\subsection{Two-dimensional POD}
\label{subsec:2dpod}
Two-dimensional POD, considering all three velocity components, is employed as
a first step in the analysis of swirl-switching.
Instantaneous velocity fields are saved at a distance of
$2D$ from the end of the bent section and are used, with their mirror
images, to assemble the snapshot matrices \citep{Berkooz1993}.
$1\,234$ velocity fields were saved at a sampling frequency of $St=0.25$, and
the sampling was started only after the solution had reached a statistically
steady state.
As a consequence of exploiting the mirror symmetry, all modes are either
symmetric or antisymmetric, a condition to which they would converge provided
that a sufficient number of snapshots had been saved.

\begin{figure}
	\centering
	\includegraphics{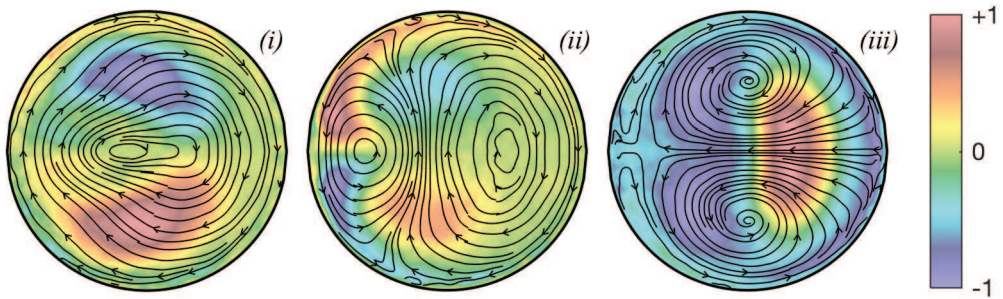}
	\caption{%
		Pseudocolor of the streamwise velocity component and streamlines of the
		in-plane velocity components for the first three POD modes
		(\textit{i}--\textit{iii}).
		The modes are oriented as in figure~\ref{fig:domain_sketch}(b).
   	The snapshots were extracted at $s_o=2D$.
	}
	\label{fig:modes_2d}
\end{figure}

\begin{figure}
	\centering
	\includegraphics[width=.8\textwidth]{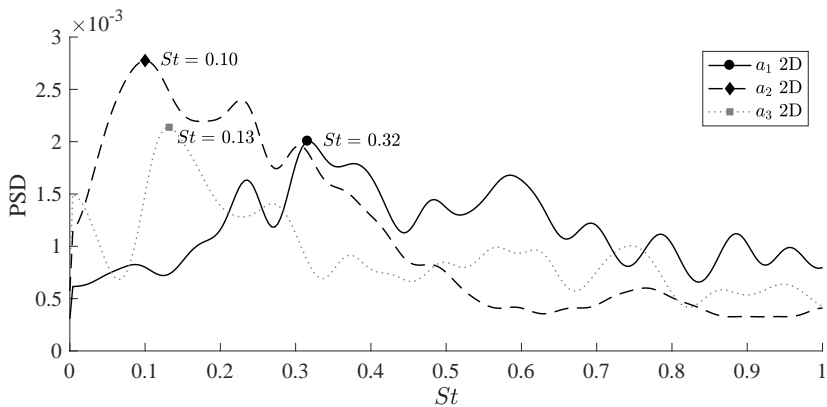}
	\caption{%
		Welch's power spectral density estimate for the time coefficients $a_i$ of the
		most energetic 2D POD modes.
		The frequencies are scaled with pipe diameter and bulk velocity.
		The 2D modes were extracted at $s_o=2D$.
		The markers and corresponding labels report the frequency of the highest
		peak of each spectrum.
	}
	\label{fig:psd_2d}
\end{figure}

The first three modes are shown in
figure~\ref{fig:modes_2d} by means of pseudocolors of their streamwise velocity
component and streamlines of the in-plane velocity components.
Two out of three modes are antisymmetric: \textit{(i, ii)} and are in the form
of a single swirl covering the whole pipe section,
\textit{(i)}, and a double swirl,
\textit{(ii)}, formed by two counter-rotating vortices disposed 
along the inner-outer direction
on the symmetry plane.
The third mode, \textit{(iii)}, resembles a harmonic of the Dean vortices.

These findings are in agreement with previous experimental work, such as that
of \citet{Sakakibara2010} and \citet{KalpakliVester2013}, which attributed the
dynamics of swirl-switching to the antisymmetric modes.
The frequency content of these modes is presented in
figure~\ref{fig:psd_2d}, in terms of Welch's power spectral density estimate for
the time coefficients of the first three modes, corresponding to the structures
shown in figure~\ref{fig:modes_2d}.
It can be observed that the spectra have a low peak-to-noise ratio and that
each mode is characterised by a different spectrum and peak frequency,
in agreement with previous 2D POD studies:
see, \textit{e.g.}, figure~8 in \citet{Hellstrom2013}, which presents peaks with
similar values to the present ones, although their study was
for a slightly larger curvature of $\delta=0.5$.
This fact has caused some confusion in the past, with disagreeing
authors attributing different causes to the various peaks, without being able
to come to the same conclusion about the frequency, nor the structure, of
swirl-switching.
The reason is that swirl-switching is caused by a three-dimensional
wave-like structure, as will be shown by 3D POD in
\S\ref{subsec:3dpod}, and a two-dimensional
cross-flow analysis cannot distinguish between the spatial and temporal
amplitude modulations created by the passage of the wave.
A simple analytical demonstration of this concept is provided in the Appendix,
and shows that conclusions drawn from a flow reconstruction based on
2D POD modes \citep[see, \textit{e.g.},][]{Hellstrom2013} are incomplete.

%-------------------------------------------------------------------------------
\subsection{Three-dimensional POD}
\label{subsec:3dpod}
For the 3D POD, the same snapshots as for 2D POD were used.
In order to reduce memory requirements, the snapshots were interpolated on a
coarser mesh before computing the POD.
This is, however, not a problem because the swirl-switching is related to
large-scale fluctuations in the flow.

\begin{figure}
	\centering
	\includegraphics[width=0.9\textwidth]{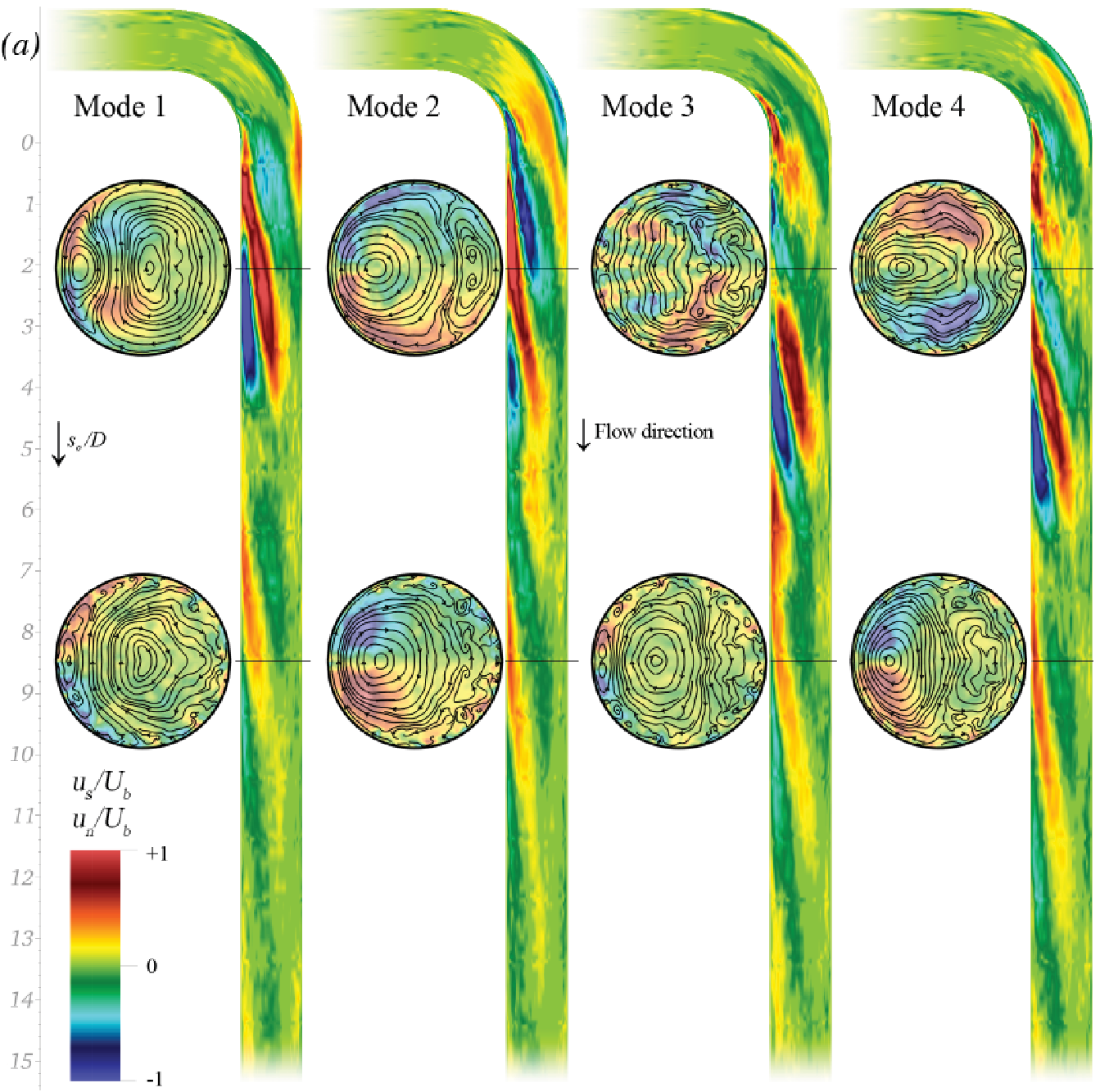}\\[5pt]
	\includegraphics[width=\textwidth]{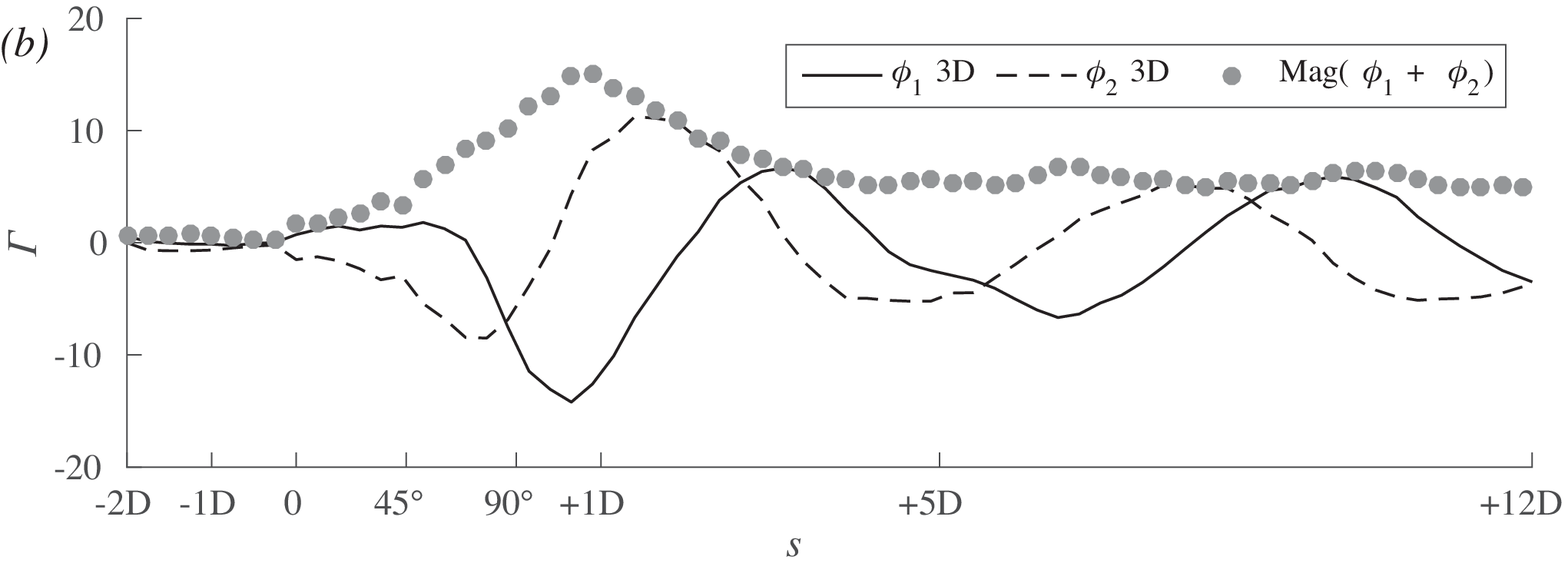}
	\caption{%
		Panel \textit{(a)} shows the
   	four most energetic three-dimensional POD modes.
		The four longitudinal cuts show pseudocolours of the normal velocity
		component $u_n$, while the eight cross-sections display the in-plane
		streamlines and are coloured by streamwise velocity $u_s$.
		The supplementary material includes two videos showing the reconstruction
		of the flow based on these modes.
		Panel \textit{(b)} shows the
		swirl intensity, measured by circulation $\Gamma$, along the streamwise
		axis of the two most energetic modes, $\phi_1$ and $\phi_2$, and their
		envelope. The spatially decaying, wave-like behaviour can be appreciated.
	}
	\label{fig:modi_d03}
\end{figure}

\begin{figure}
	\centering
	\includegraphics[width=0.66\textwidth]{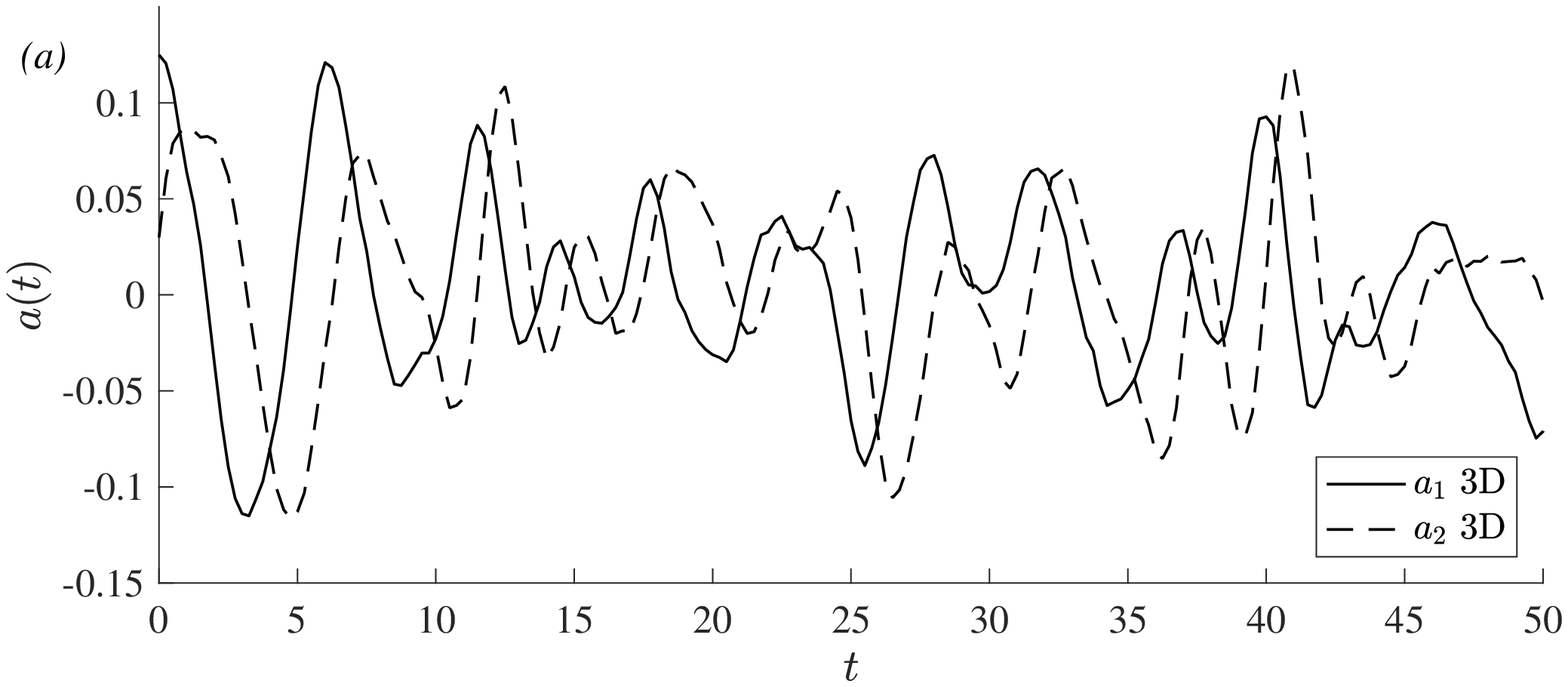}
	\includegraphics[width=0.33\textwidth]{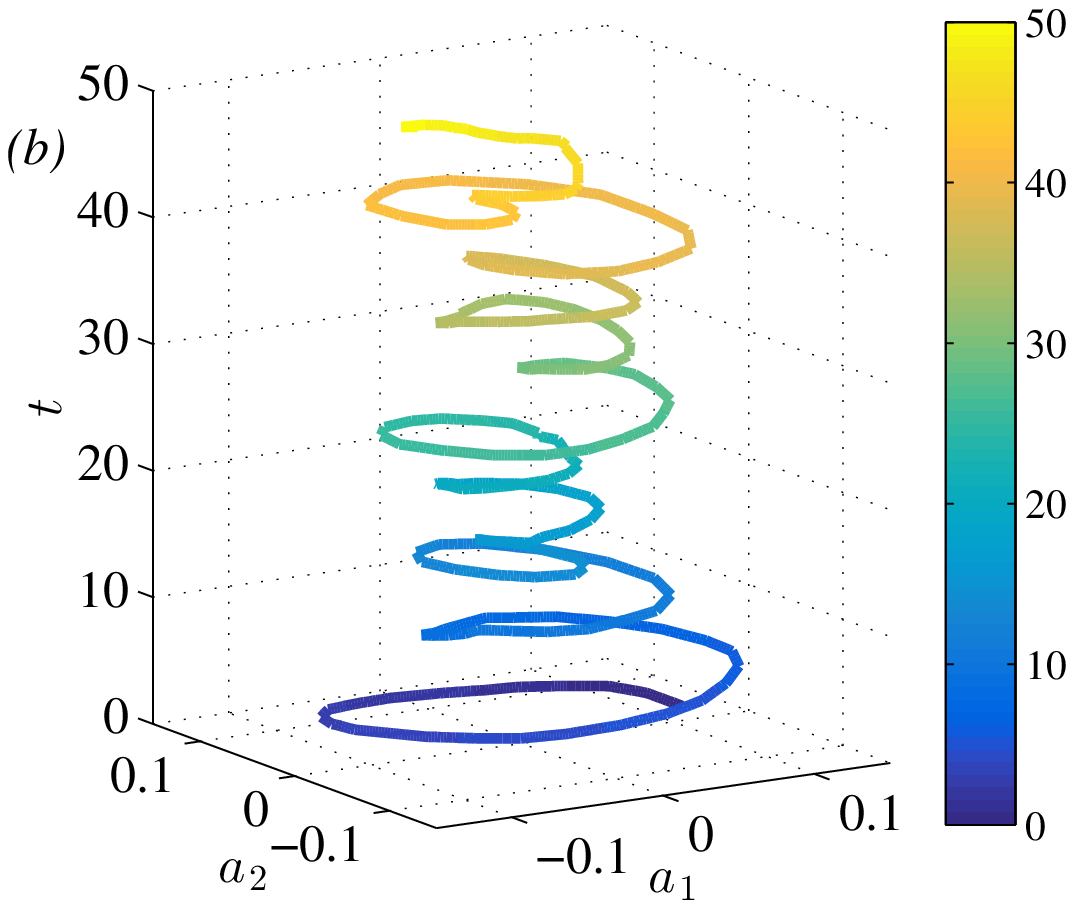}
	\caption{%
		Time coefficients $a_i(t)$ of the two most energetic three-dimensional
		POD modes.
		Panel \textit{(a)} shows the temporal signal, which allows to
		observe the qualitative quarter-period phase shift of mode 2 with respect
		to mode 1;
		panel \textit{(b)} shows the (colorcoded) time over coefficients $a_1$
		and $a_2$, illustrating the oscillating character.
		The time axis is (arbitrarily) cut at $t = 50 D/U_b$ for illustration
		purposes, the total recorded signal is over $300 D/U_b$.
	}
	\label{fig:tcoeffs_d03}
\end{figure}

\begin{figure}
	\centering
	\includegraphics[width=.8\textwidth]{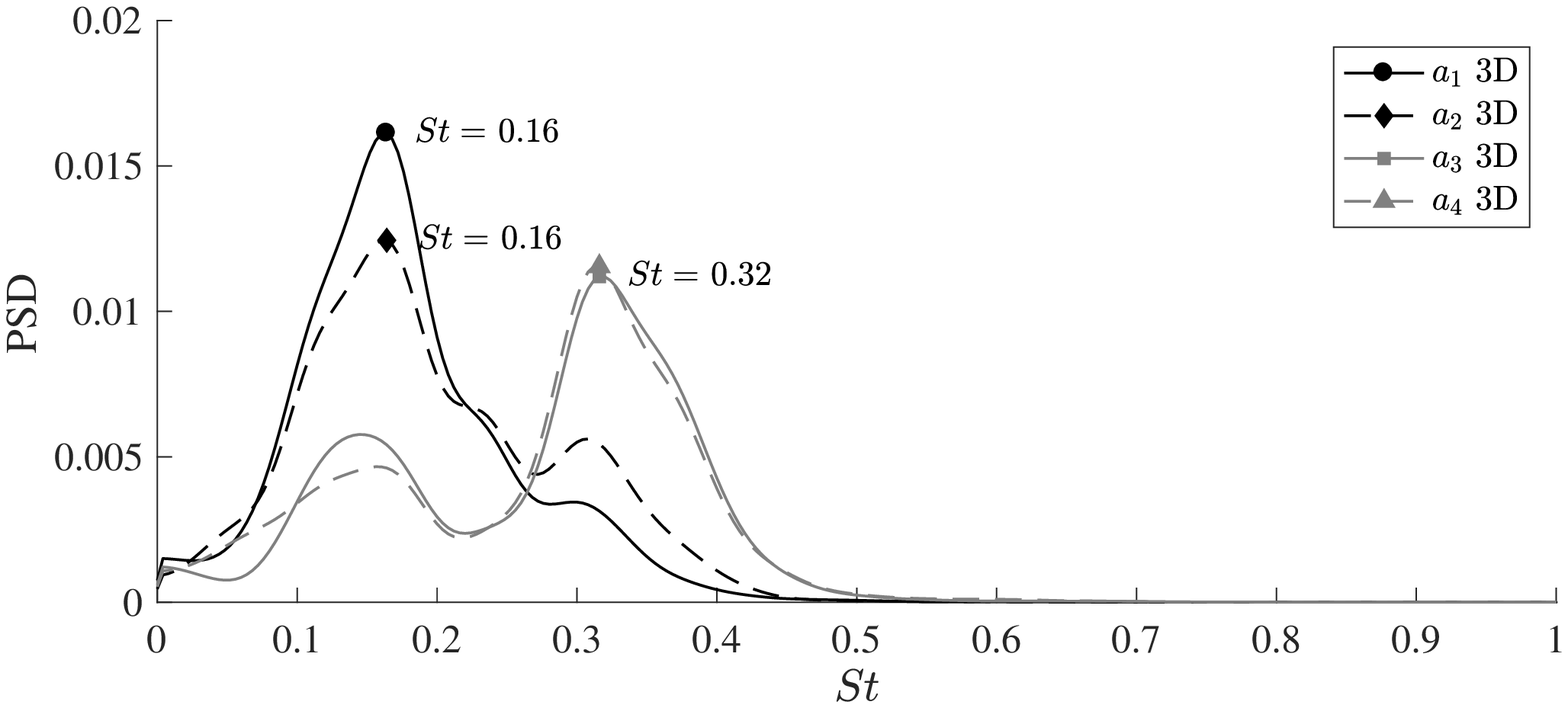}
	\caption{%
		Welch's power spectral density estimate for the time coefficients $a_i$
		of the most energetic 3D POD modes.
		The frequencies are scaled with pipe diameter and bulk velocity.
		The markers and corresponding labels report the frequency corresponding
		to the peak of each spectrum.
		The range of the Strouhal number is identical to that of
		figure~\ref{fig:psd_2d} to ease comparison.
	}
	\label{fig:psd_3d}
\end{figure}

The four most energetic modes are depicted in
figure~\ref{fig:modi_d03} by means of pseudocolors of normal and streamwise
velocity components, as well as streamlines of the in-plane velocity.
It can be observed that the modes come in pairs: 1-2 and 3-4, as is usual for
POD modes and their time coefficients in a convective flow.
The first coherent structure extracted by the POD is formed by modes $1$ and $2$
and constitutes a damped wave-like structure that is convected by the mean flow
(see figure~\ref{fig:modi_d03} for the spatial structure and
figure~\ref{fig:tcoeffs_d03} for the corresponding time coefficients;
the supplementary video \texttt{movie2\_modes0-2.mov} shows their behaviour in
time).
This is not a travelling wave such as those observed in transitional flows,
as the ones of the examples mentioned in the introduction, but a coherent structure
extracted by POD from a developed turbulent background that persists in fully
developed turbulence, and is just a regular component of the flow on which
irregular turbulent fluctuations are superimposed
\citep[see, \textit{e.g.},][for a similar case]{Manhart1993}.
Nevertheless, the present wave-like structure could be a surviving
remnant of pre-existing, purely time-periodic, flow structures formed in the bent section and arising in the
process of transition to turbulence past bends
\citep[see, \textit{e.g.}, the case of the flow past a circular cylinder by][]{sipp-lebedev-2007}.
It was found that the first instability of the flow inside of a toroidal pipe
is characterised by the appearance of travelling waves
\citep[][]{Kuhnen2014,Canton2016a}.
It is therefore possible that similar waves appear in the transition to
turbulence of the present flow case, and continue to modulate the large scales
of the flow at high Reynolds numbers while being submerged in small-scale
turbulence.
To support this hypothesis, the frequencies and wavelengths of the present
coherent structures are in the same range as those measured in toroidal pipes
\citep[][]{Canton2016a}, and \citet{Brucker1998} observed swirl-switching even
for $\Rey$ as low as $2\,000$, although the measured oscillations had very low
amplitude.

The present modes are, obviously, not strictly periodic in space nor in time:
as can be seen in figure~\ref{fig:modi_d03}\textit{(b)} showing the
swirl intensity,
the intensity of the modes
is essentially zero upstream of the bend ($s<0$), reaches a maximum at about
$1D$ downstream of the bend end, and then
decreases with the distance from the bend.
Furthermore, the respective
time coefficients are only quasiperiodic, as
can be observed from their temporal signal, depicted in
figure~\ref{fig:tcoeffs_d03}\textit{(a)}, and by their frequency spectra,
figure~\ref{fig:psd_3d}\textit{(a)}.
Nevertheless, it can be observed in figure~\ref{fig:modi_d03} that the spatial
structure of these modes is qualitatively sinusoidal along the streamwise
direction $s_o$, with a wavelength of about $7$ pipe diameters.
The figures in \citet{Brucker1998} actually already suggest the appearance of a
wave-like structure in the presence of swirl-switching.

This wave-like structure is formed by two counter-rotating swirls,
visible in the 2D cross-sections in figure~\ref{fig:modi_d03}, which are
advected in the streamwise direction while decaying in intensity and, at the
same time, move from the inside of the bend towards the outside, as can be seen
in the longitudinal cuts in figure~\ref{fig:modi_d03} and in the supplementary
video \texttt{movie2\_modes0-2.mov}.
The temporal amplitude of these modes is also qualitatively cyclic, as
illustrated by the projection along the time coefficients in
figure~\ref{fig:tcoeffs_d03}\textit{(b)}.
The wave-like behaviour can be appreciated even better in the
aforementioned video showing the flow reconstructed with these two modes,
\texttt{movie2\_modes0-2.mov}.
Modes $1$ and $2$ are phase-shifted by a quarter of their quasi-period:
figure~\ref{fig:modi_d03} shows that the structure of mode $2$ is located
approximately a quarter of a wavelength further downstream of the structure of
mode $1$;
while, figure~\ref{fig:tcoeffs_d03} illustrates the constant delay of the time
coefficient of mode $2$ with respect to that of mode $1$.

The second structure, formed by modes $3$ and $4$, has a spatial layout
that closely resembles that of the first pair, \ie it is also a
wave-like structure, and constitutes the first
``harmonic'' of the wave formed by modes $1$ and $2$.
The spatial structure of modes $3$ and $4$ has half
of the main wavelength of modes $1$ and $2$, and
the highest peak in the spectrum of the third and fourth time coefficients
is at exactly twice the frequency of the peaks of $a_1$ and $a_2$, as can
be seen in figure~\ref{fig:psd_3d}\textit{(a)}.
The video \texttt{movie3\_modes0-4.mov} shows the reconstruction of the flow
field by including modes $3$ and $4$.
It can be observed that these modes introduce oscillations with higher
frequency and smaller amplitude when compared to the reconstruction employing
only modes $1$ and $2$.

As can be observed from figure~\ref{fig:modi_d03}, the modes do not present
any connection to the straight pipe section preceding the bend.
This is in direct contrast with the findings of \citet{Carlsson2015}, whose
results were likely altered by the interference of an intrinsic frequency and
wavelength on the recycling inflow boundary with the structure of the
swirl-switching.
Our results are, instead, in agreement with \citet{Noorani2016} who
observed swirl-switching in a toroidal pipe (\ie in the absence of a straight
upstream section) confirming that these large-scale oscillations are not
caused by structures formed in the straight pipe, but by an effect which is
intrinsic to the bent section.

The power spectral density analysis of the time coefficients of these
modes, computed as a Welch's estimate, is presented in figure~\ref{fig:psd_3d}.
One can see that, unlike the PSD of the two-dimensional POD modes
(figure~\ref{fig:psd_2d}), the three-dimensional modes present two
distinct peaks, one per pair of modes.
The peak for the first modal pair is located at $St \approx 0.16$, which is
in the range of Strouhal numbers found by both \citet{Brucker1998} and
\citet{Hellstrom2013}.
More importantly, this frequency is the lowest for this pair of modes and
matches that given by the wavelength and propagation speed of the wave
as well as that of the swirl-switching, as observed by reconstructing the flow
field with the most energetic POD modes (see the online movies).

The present analyses were also performed on a pipe with curvature
$\delta=0.1$ for the same Reynolds number.
Swirl-switching was observed in this case as well, with dynamics which is
qualitatively identical to the one observed for $\delta=0.3$, but is
characterised by lower frequencies, peaking at $St\approx0.045$.
The lower frequencies and larger scales (wavelength of about $20D$)
characterising the wave-like structure at this curvature meant that
a quantitative analysis was too expensive with the present setup.
We have therefore limited this work to the study of one curvature only, but
preliminary, not converged results can be found in \citet{Hufnagel2016}.

%-------------------------------------------------------------------------------
\section{Summary and conclusions}
\label{sec:conclusions}
This work presents the first DNS analysis of swirl-switching in a $90^\circ$
bent pipe.
The simulations were performed by using a synthetic eddy method to generate
high-quality inflow conditions, in an effort to avoid any interference
between the incoming flow and the dynamics of the flow in the bent section, as
was observed in previous studies.
Three-dimensional POD was used to isolate the dominant structures of the flow.
This method allowed the identification of a wave-like
structure, originating in the bent section, constituted by the first modal
pair.
A reconstruction of the flow field using the most energetic modal pair
confirmed that the swirl-switching is caused by this structure.

The swirl-switching frequency found in the present study is in the range of
those deduced by \citet{Brucker1998} and \citet{Hellstrom2013}.
The structure of the modes, which presents no connection to the
upstream straight pipe, confirms what
was conjectured by \citet{Noorani2016}, who observed swirl-switching in a
toroidal pipe, namely that swirl-switching is a phenomenon intrinsic to
the bent pipe section.

Clearly, the present findings are in contrast with previous conclusions drawn 
from flow reconstructions based on 2D POD modes and Taylor's frozen turbulence 
hypothesis \citep[see, \textit{e.g.},][]{Hellstrom2013}:
the 2D analysis mixes convection and true temporal variation, and thus cannot 
reveal the full three-dimensional structure of travelling modes.
This does not only apply to the present flow case, but to any streamwise
inhomogeneous flow in which 2D POD is utilised in the cross-flow direction.

The wave-like structure found in the present study is different from
those observed in transitional flows \citep[see, \textit{e.g.}][]{Hof2004},
in the sense that it is simply a coherent structure extracted by POD from a
turbulent background flow, as opposed to an exact coherent state.
Nevertheless, we conjecture that this structure may be a surviving remnant 
of a global instability caused by the bend \citep{Kuhnen2014,Canton2016a}.

\vspace{\baselineskip}

Financial support by the Swedish Research Council (VR) is gratefully acknowledged.
Computer time was provided by the Swedish National Infrastructure for Computing (SNIC).
We acknowledge that part of the results of this research have been achieved using the DECI resource SiSu based in Finland at CSC with support from the PRACE aisbl.
This material is also based in part upon work supported by the US Department of
Energy, office of Science, under contract DE-AC02-06CH11357.

%===============================================================================
\oneappendix
\section{Considerations on the use of 2D POD}
\label{subsec:2d_pod_fail}
This section explains, analytically, the reasons why a two-dimensional
cross-flow POD analysis is an ineffective tool for understanding
swirl-switching.
In order to capture the essence of the phenomenon, the example is without
spatial dissipation and noise, but these can be added at will without changing
the discussion or the results.
A Matlab script performing the operations described in this section is
provided as part of the supplementary online material.

Consider a sine wave of period $2\pi l$, travelling at speed $v$, and with
amplitude modulated at a frequency $\omega/(2\pi)$:
\begin{equation}
	g(x,t) = \sin\left(\frac{x - vt}{l}\right) \cos\left(\omega t\right).
\end{equation}
When measuring its passage at a given spatial position, say $g(x_m, t)$, the
recorded time signal will contain two frequencies,
$f_1 = (\omega-v/l)/ (2\pi)$ and
$f_2 = (\omega+v/l)/ (2\pi)$,
that combine the spatial component, $f_s = v/(2\pi l)$, and the temporal
component, $f_t = \omega/(2\pi)$.
This combination is a result of the fact that $g$ can be rewritten, using
one prosthaphaeresis formula, separating the time and space dependencies:
\begin{equation}
	g(x,t) = \frac{1}{2}\sin\left[\frac{x}{l} + \left(\omega-\frac{v}{l}\right)t\right]
	       + \frac{1}{2}\sin\left[\frac{x}{l} - \left(\omega+\frac{v}{l}\right)t\right].
\end{equation}
The two components, $f_s$ and $f_t$, would be measured in isolation if the
function $g$ were a pure travelling wave ($\omega=0$) or a pure standing wave
($v=0$).
However, when both aspects are present ($\omega\neq0$ and $v\neq0$) a complete
knowledge of $g$ is necessary in order to separate $f_s$ from $f_t$.
This, clearly, is possible in the present example, where the analytical
expression of $g(x,t)$ is known.
When studying an unknown phenomenon (such as swirl-switching) the knowledge of
$f_1$ and $f_2$ is insufficient: one does not know what is causing the measured
frequencies: it could be two travelling waves advected at different speeds (or
provided with different period); two standing waves modulated at different
frequencies; or, as in this case, one travelling wave with modulated amplitude.

This problem can transferred to a POD analysis as well:
the 2D POD in the pipe corresponds to a zero-dimensional POD in this example,
which employs the measurements $g(x_m,t_i)$ as snapshots,
while the 3D POD of the bent pipe flow corresponds to a one-dimensional POD
which uses the function $g(x,t_i)$ over the whole $x$ domain as snapshots.

The 0D POD returns a single mode which assumes a value of either $+1$ or
$-1$ and does not provide any information about the spatial structure of $g$.
The spectrum of the time coefficient corresponding to this single mode contains
both frequencies $f_1$ and $f_2$.
When using 0D POD one does not have any information about the spatial nature of
$g$, and is lead to believe that the oscillations measured in $x_m$ are
caused by two periodic phenomena with frequencies $f_1$ and $f_2$.
This likely is what has caused so much disagreement in the literature about the value
of the Strouhal number related to the swirl-switching and on the 2D POD mode
responsible for this phenomenon.
The answer is that none of the 2D POD modes reported in the literature is
actually the swirl-switching mode, and the Strouhal numbers extracted from time
coefficients do not provide a correct description.

A 1D POD analysis of the function $g$, which is the analogue of the 3D POD in
the bent pipe, provides the correct answers.
It results in two sinusoidal modes which, with the corresponding time
coefficients, reproduce the complete travelling and oscillatory behaviour of
$g$.
The spectra of the time coefficients still contain only $f_1$ and $f_2$, but
have a much higher peak to noise ratio compared to the 0D POD, as observed in
the bent pipe by comparing figures \ref{fig:psd_2d} and \ref{fig:psd_3d}.
Moreover, by analysing the reconstruction of $g$, they allow the separation of
$f_s$ from $f_t$.

It is now clear why in the case of a streamwise-dependent spatial structure,
such as the one creating swirl-switching (as shown in \S\ref{subsec:3dpod}),
only a fully three-dimensional analysis can correctly identify the
actual spatial and temporal components.

%===============================================================================
\bibliographystyle{jfm}
\bibliography{references}

\begin{thebibliography}{37}
\expandafter\ifx\csname natexlab\endcsname\relax\def\natexlab#1{#1}\fi
\def\au#1{#1} \def\ed#1{#1} \def\yr#1{#1}\def\at#1{#1}\def\jt#1{\textit{#1}}
  \def\bt#1{#1}\def\bvol#1{\textbf{#1}} \def\vol#1{#1} \def\pg#1{#1}
  \def\publ#1{#1}\def\arxiv#1{#1}\def\org#1{#1}\def\st#1{\textit{#1}}

\bibitem[Anwer {\em et~al.\/}(1989)Anwer, So \& Lai]{Anwer1989}
{\sc \au{Anwer, M.}, \au{So, R. M.~C.} \& \au{Lai, Y.~G.}} \yr{1989}
  \at{{Perturbation by and recovery from bend curvature of a fully developed
  turbulent pipe flow}}.  \jt{Phys. Fluids}  \bvol{1},  \pg{1387--1397}.

\bibitem[Berkooz {\em et~al.\/}(1993)Berkooz, Holmes \& Lumley]{Berkooz1993}
{\sc \au{Berkooz, G.}, \au{Holmes, P.} \& \au{Lumley, J.~L.}} \yr{1993}
  \at{{The proper orthogonal decomposition in the analysis of turbulent
  flows}}.  \jt{Annu. Rev. Fluid Mech.}  \bvol{25},  \pg{539--575}.

\bibitem[Boussinesq(1868)]{Boussinesq1868}
{\sc \au{Boussinesq, M.~J.}} \yr{1868}  \at{{M{\'{e}}moire sur l'influence des
  frottements dans les mouvements r{\'{e}}guliers des fluides}}.  \jt{J. Math.
  Pure Appl.}  \bvol{13},  \pg{377--424}.

\bibitem[Br{\"{u}}cker(1998)]{Brucker1998}
{\sc \au{Br{\"{u}}cker, C.}} \yr{1998} {A time-recording DPIV-study of the
  swirl-switching effect in a $90^\circ$ bend flow}.  \bt{In {\em Proc. 8th
  Int. Symp. Flow Vis.\/}},  \pg{pp. 171.1--171.6}. Sorrento (NA), Italy.

\bibitem[Canton {\em et~al.\/}(2017)Canton, {\"O}rl{\"u} \&
  Schlatter]{Canton2016b}
{\sc \au{Canton, J.}, \au{{\"O}rl{\"u}, R.} \& \au{Schlatter, P.}} \yr{2017}
  \at{Characterisation of the steady, laminar incompressible flow in toroidal
  pipes covering the entire curvature range}.  \jt{Int. J. Heat Fluid Flow}
  \bvol{66},  \pg{95--107}.

\bibitem[Canton {\em et~al.\/}(2016)Canton, Schlatter \&
  {\"{O}}rl{\"{u}}]{Canton2016a}
{\sc \au{Canton, J.}, \au{Schlatter, P.} \& \au{{\"{O}}rl{\"{u}}, R.}}
  \yr{2016}  \at{{Modal instability of the flow in a toroidal pipe}}.  \jt{J.
  Fluid Mech.}  \bvol{792},  \pg{894--909}.

\bibitem[Carlsson {\em et~al.\/}(2015)Carlsson, Alenius \& Fuchs]{Carlsson2015}
{\sc \au{Carlsson, C.}, \au{Alenius, E.} \& \au{Fuchs, L.}} \yr{2015}
  \at{{Swirl switching in turbulent flow through $90^\circ$ pipe bends}}.
  \jt{Phys. Fluids}  \bvol{27},  \pg{085112}.

\bibitem[Chandler \& Northrop(2003)]{Chandler2003}
{\sc \au{Chandler, R.} \& \au{Northrop, P.}} \yr{2003} {Fortran random number
  generation}. {\url{http://www.ucl.ac.uk/~ucakarc/work/randgen.html}}.

\bibitem[Dean(1928)]{Dean1928a}
{\sc \au{Dean, W.~R.}} \yr{1928}  \at{{The streamline motion of fluid in a
  curved pipe}}.  \jt{Phil. Mag.}  \bvol{5},  \pg{673--693}.

\bibitem[Doherty {\em et~al.\/}(2007)Doherty, Monty \& Chong]{Doherty2007}
{\sc \au{Doherty, J.}, \au{Monty, J.} \& \au{Chong, M.}} \yr{2007} {The
  development of turbulent pipe flow}.  \bt{In {\em 16th Australas. Fluid Mech.
  Conf.\/}},  \pg{pp. 266--270}.

\bibitem[{El Khoury} {\em et~al.\/}(2013){El Khoury}, Schlatter, Noorani,
  Fischer, Brethouwer \& Johansson]{ElKhoury2013}
{\sc \au{{El Khoury}, G.~K.}, \au{Schlatter, P.}, \au{Noorani, A.},
  \au{Fischer, P.~F.}, \au{Brethouwer, G.} \& \au{Johansson, A.~V.}} \yr{2013}
  \at{{Direct numerical simulation of turbulent pipe flow at moderately high
  Reynolds numbers}}.  \jt{Flow, Turbul. Combust.}  \bvol{91},  \pg{475--495}.

\bibitem[Eustice(1910)]{Eustice1910}
{\sc \au{Eustice, J.}} \yr{1910}  \at{{Flow of water in curved pipes}}.
  \jt{Proc. R. Soc. London, Ser. A}  \bvol{84},  \pg{107--118}.

\bibitem[Fischer {\em et~al.\/}(2008)Fischer, Lottes \& Kerkemeier]{nek5000}
{\sc \au{Fischer, P.~F.}, \au{Lottes, J.~W.} \& \au{Kerkemeier, S.~G.}}
  \yr{2008} {Nek5000 Web page, http://nek5000.mcs.anl.gov}.

\bibitem[Hellstr{\"{o}}m {\em et~al.\/}(2013)Hellstr{\"{o}}m, Zlatinov, Cao \&
  Smits]{Hellstrom2013}
{\sc \au{Hellstr{\"{o}}m, L. H.~O.}, \au{Zlatinov, M.~B.}, \au{Cao, G.} \&
  \au{Smits, A.~J.}} \yr{2013}  \at{{Turbulent pipe flow downstream of a
  $90^\circ$ bend}}.  \jt{J. Fluid Mech.}  \bvol{735},  \pg{R7}.

\bibitem[Hof {\em et~al.\/}(2004)Hof, van Doorne, Westerweel, Nieuwstadt,
  Faisst, Eckhardt, Wedin, Kerswell \& Waleffe]{Hof2004}
{\sc \au{Hof, B.}, \au{van Doorne, C. W.~H.}, \au{Westerweel, J.},
  \au{Nieuwstadt, F. T.~M.}, \au{Faisst, H.}, \au{Eckhardt, B.}, \au{Wedin,
  H.}, \au{Kerswell, R.~R.} \& \au{Waleffe, F.}} \yr{2004}  \at{{Experimental
  observation of nonlinear traveling waves in turbulent pipe flow}}.
  \jt{Science}  \bvol{305},  \pg{1594--1598}.

\bibitem[Hufnagel(2016)]{Hufnagel2016}
{\sc \au{Hufnagel, L.}} \yr{2016}  \at{{On the swirl-switching in developing
  bent pipe flow with direct numerical simulation}}. Msc thesis, KTH Mechanics,
  Stockholm, Sweden.

\bibitem[Jarrin {\em et~al.\/}(2006)Jarrin, Benhamadouche, Laurence \&
  Prosser]{Jarrin2006}
{\sc \au{Jarrin, N.}, \au{Benhamadouche, S.}, \au{Laurence, D.} \& \au{Prosser,
  R.}} \yr{2006}  \at{{A synthetic-eddy-method for generating inflow conditions
  for large-eddy simulations}}.  \jt{Int. J. Heat Fluid Flow}  \bvol{27},
  \pg{585--593}.

\bibitem[{Kalpakli} \& {\"{O}}rl{\"{u}}(2013)]{KalpakliVester2013}
{\sc \au{{Kalpakli}, A.} \& \au{{\"{O}}rl{\"{u}}, R.}} \yr{2013}
  \at{{Turbulent pipe flow downstream a $90^\circ$ pipe bend with and without
  superimposed swirl}}.  \jt{Int. J. Heat Fluid Flow}  \bvol{41},
  \pg{103--111}.

\bibitem[{Kalpakli Vester} {\em et~al.\/}(2015){Kalpakli Vester},
  {\"{O}}rl{\"{u}} \& Alfredsson]{KalpakliVester2015}
{\sc \au{{Kalpakli Vester}, A.}, \au{{\"{O}}rl{\"{u}}, R.} \& \au{Alfredsson,
  P.~H.}} \yr{2015}  \at{{POD analysis of the turbulent flow downstream a mild
  and sharp bend}}.  \jt{Exp. Fluids}  \bvol{56},  \pg{57}.

\bibitem[{Kalpakli Vester} {\em et~al.\/}(2016){Kalpakli Vester},
  {\"{O}}rl{\"{u}} \& Alfredsson]{KalpakliVester2016}
{\sc \au{{Kalpakli Vester}, A.}, \au{{\"{O}}rl{\"{u}}, R.} \& \au{Alfredsson,
  P.~H.}} \yr{2016}  \at{{Turbulent flows in curved pipes: recent advances in
  experiments and simulations}}.  \jt{Appl. Mech. Rev.}  \bvol{68},
  \pg{050802}.

\bibitem[K{\"{u}}hnen {\em et~al.\/}(2015)K{\"{u}}hnen, Braunshier, Schwegel,
  Kuhlmann \& Hof]{Kuhnen2015}
{\sc \au{K{\"{u}}hnen, J.}, \au{Braunshier, P.}, \au{Schwegel, M.},
  \au{Kuhlmann, H.~C.} \& \au{Hof, B.}} \yr{2015}  \at{{Subcritical versus
  supercritical transition to turbulence in curved pipes}}.  \jt{J. Fluid
  Mech.}  \bvol{770},  \pg{R3}.

\bibitem[K{\"{u}}hnen {\em et~al.\/}(2014)K{\"{u}}hnen, Holzner, Hof \&
  Kuhlmann]{Kuhnen2014}
{\sc \au{K{\"{u}}hnen, J.}, \au{Holzner, M.}, \au{Hof, B.} \& \au{Kuhlmann,
  H.~C.}} \yr{2014}  \at{{Experimental investigation of transitional flow in a
  toroidal pipe}}.  \jt{J. Fluid Mech.}  \bvol{738},  \pg{463--491}.

\bibitem[Lumley(1967)]{Lumley1967}
{\sc \au{Lumley, J.~L.}} \yr{1967}  \at{{The structure of inhomogeneous
  turbulent flows}}.  \bt{In {\em Atmos. Turbul. Radio Wave Propag.\/} (ed.
  \ed{A.~M. Yaglom \& V.~I. Tatarski})},  \pg{pp. 166--178}.  \publ{Moscow}.

\bibitem[Manhart \& Wengle(1993)]{Manhart1993}
{\sc \au{Manhart, M.} \& \au{Wengle, H.}} \yr{1993}  \at{{A spatiotemporal
  decomposition of a fully inhomogeneous turbulent flow field}}.  \jt{Theor.
  Comput. Fluid Dyn.}  \bvol{5},  \pg{223--242}.

\bibitem[Noorani {\em et~al.\/}(2013)Noorani, {El Khoury} \&
  Schlatter]{Noorani2013}
{\sc \au{Noorani, A.}, \au{{El Khoury}, G.~K.} \& \au{Schlatter, P.}} \yr{2013}
   \at{{Evolution of turbulence characteristics from straight to curved
  pipes}}.  \jt{Int. J. Heat Fluid Flow}  \bvol{41},  \pg{16--26}.

\bibitem[Noorani \& Schlatter(2016)]{Noorani2016}
{\sc \au{Noorani, A.} \& \au{Schlatter, P.}} \yr{2016}  \at{{Swirl-switching
  phenomenon in turbulent flow through toroidal pipes}}.  \jt{Int. J. Heat
  Fluid Flow}  \bvol{61},  \pg{108--116}.

\bibitem[Poletto {\em et~al.\/}(2013)Poletto, Craft \& Revell]{Poletto2013}
{\sc \au{Poletto, R.}, \au{Craft, T.} \& \au{Revell, A.}} \yr{2013}  \at{{A new
  divergence free synthetic eddy method for the reproduction of inlet flow
  conditions for LES}}.  \jt{Flow, Turbul. Combust.}  \bvol{91},
  \pg{519--539}.

\bibitem[Poletto {\em et~al.\/}(2011)Poletto, Revell, Craft \&
  Jarrin]{Poletto2011}
{\sc \au{Poletto, R.}, \au{Revell, A.}, \au{Craft, T.~J.} \& \au{Jarrin, N.}}
  \yr{2011} {Divergence free synthetic eddy method for embedded LES inflow
  boundary conditions}.  \bt{In {\em 7th Int. Symp. Turbul. Shear Flow
  Phenom.\/}}. Ottawa.

\bibitem[R{\"{u}}tten {\em et~al.\/}(2001)R{\"{u}}tten, Meinke \&
  Schr{\"{o}}der]{Rutten2001}
{\sc \au{R{\"{u}}tten, F.}, \au{Meinke, M.} \& \au{Schr{\"{o}}der, W.}}
  \yr{2001}  \at{{Large-eddy simulations of $90^\circ$ pipe bend flows}}.
  \jt{J. Turbul.}  \bvol{2},  \pg{N3}.

\bibitem[R{\"{u}}tten {\em et~al.\/}(2005)R{\"{u}}tten, Schr{\"{o}}der \&
  Meinke]{Rutten2005}
{\sc \au{R{\"{u}}tten, F.}, \au{Schr{\"{o}}der, W.} \& \au{Meinke, M.}}
  \yr{2005}  \at{{Large-eddy simulation of low frequency oscillations of the
  Dean vortices in turbulent pipe bend flows}}.  \jt{Phys. Fluids}  \bvol{17},
  \pg{035107}.

\bibitem[Sakakibara \& Machida(2012)]{Sakakibara2012}
{\sc \au{Sakakibara, J.} \& \au{Machida, N.}} \yr{2012}  \at{{Measurement of
  turbulent flow upstream and downstream of a circular pipe bend}}.  \jt{Phys.
  Fluids}  \bvol{24},  \pg{041702}.

\bibitem[Sakakibara {\em et~al.\/}(2010)Sakakibara, Sonobe, Goto, Tezuka, Tada
  \& Tezuka]{Sakakibara2010}
{\sc \au{Sakakibara, J.}, \au{Sonobe, R.}, \au{Goto, H.}, \au{Tezuka, H.},
  \au{Tada, H.} \& \au{Tezuka, K.}} \yr{2010} {Stereo-PIV study of turbulent
  flow downstream of a bend in a round pipe}.  \bt{In {\em 14th Int. Symp. Flow
  Vis.\/}}. EXCO Daegu, Korea.

\bibitem[Sipp \& Lebedev(2007)]{sipp-lebedev-2007}
{\sc \au{Sipp, D.} \& \au{Lebedev, A.}} \yr{2007}  \at{Global stability of base
  and mean flows: a general approach and its applications to cylinder and open
  cavity flows}.  \jt{J. Fluid Mech.}  \bvol{593},  \pg{333--358}.

\bibitem[Sirovich(1987)]{Sirovich1987}
{\sc \au{Sirovich, L.}} \yr{1987}  \at{{Turbulence and the dynamics of coherent
  structures. Part I: coherent structures}}.  \jt{Q. Appl. Math.}  \bvol{45},
  \pg{561--571}.

\bibitem[Sudo {\em et~al.\/}(1998)Sudo, Sumida \& Hibara]{Sudo1998}
{\sc \au{Sudo, K.}, \au{Sumida, M.} \& \au{Hibara, H.}} \yr{1998}
  \at{{Experimental investigation on turbulent flow in a circular-sectioned
  90-degree bend}}.  \jt{Exp. Fluids}  \bvol{25},  \pg{42--49}.

\bibitem[Tunstall \& Harvey(1968)]{Tunstall1968}
{\sc \au{Tunstall, M.~J.} \& \au{Harvey, J.~K.}} \yr{1968}  \at{{On the effect
  of a sharp bend in a fully developed turbulent pipe-flow}}.  \jt{J. Fluid
  Mech.}  \bvol{34},  \pg{595--608}.

\bibitem[Vashisth {\em et~al.\/}(2008)Vashisth, Kumar \& Nigam]{Vashisth2008a}
{\sc \au{Vashisth, S.}, \au{Kumar, V.} \& \au{Nigam, K. D.~P.}} \yr{2008}
  \at{{A review on the potential applications of curved geometries in process
  industry}}.  \jt{Ind. Eng. Chem. Res.}  \bvol{47},  \pg{3291--3337}.

\end{thebibliography}

\end{document}